\documentclass[prb,reprint]{revtex4-1} 

\usepackage{svg}
\usepackage{amsmath}  
\usepackage{amsfonts} 
\usepackage[utf8]{inputenc}
\usepackage{graphicx}

\begin{document}


\title{
{3D Hypersound Microscopy} of Van der Waals Heterostructures}

\author{A. Yu. Klokov}
\author{N. Yu. Frolov}
\author{A. I. Sharkov}
\author{S.N. Nikolaev}
\author{S.I. Chentsov}
\author{M.A. Chernopitssky}
\author{M.V. Pugachev}
\author{A.I. Duleba}
\author{A.V. Shupletsov}
\author{V.S. Krivobok}
\author{A. Yu. Kuntsevich}
\email{alexkun@lebedev.ru} 
\affiliation{P.N. Lebedev Physical Institute of the RAS, Leninsky Prospekt 53, Moscow, 119991, Russia}

\date{\today}

\begin{abstract}

We employ here a picosecond ultrasonic technique to study Van der Waals heterostructures. Temporal variation of the reflection coefficient of Al film that covers Van der Waals h-BN/WSe$_2$/h-BN heterostructures on a sapphire substrate after the femtosecond laser pulse excitation is carefully measured using an interferometric technique with spatial resolution. The laser pulse generates a broadband sound wave packet in aluminum film propagating perpendicular to the plane direction and partially reflecting from the heterostructural interfaces. {The demonstrated technique has enough sensitivity to resolve a WSe$_2$ monolayer embedded in hBN.}
We apply a multilayered model of the optical and acoustical response that allows to evaluate the mechanical parameters, in particular, rigidity of interfaces, inaccesiible from the other measurements.   Mapping of the Fourier spectra of the response clearly visualizes different composition regions and can therefore serve as an acoustic tomography tool.
{Our findings demonstrate almost zero acoustic phonon dissipation below 150 GHz at the interfaces and in the layers that makes Van der Waals heterostructures perspective for nano-acoustical applications.}

\end{abstract}

\maketitle 

Mechanical and acoustical properties of interlayer interfaces between two-dimensional materials essentially determine their numerous atomically thin applications.
For example transition metal dichalcogenides (TMDC) are promising candidates for future atomically thin flexible opto-electronics\cite{opto1,opto2}, atomically thin wearable thermoelectic generators\cite{wear}. Graphene demonstrates its potential for atomically thin tacktile sensors\cite{tacktile}, acoustic coatings\cite{coat}, elements of micro-\cite{mems1} and nanoelectromechanical systems\cite{nems1}. Acoustic phonons are also responsible for the heat conduction, that limits thermoelectrical\cite{thermoel1} and bolometrical\cite{bolom1,bolom2} applications of 2D materials. Importantly, maximal functionality requires assembly of Van der Waals heterostructures\cite{geim2013}, i.e. stacks of different layers, those are much more complex, compared to single layer. The performance of such structures depends on their lateral homogeneity and interfaces between layers, which can significantly impede the transport of charge carriers and phonons.
 Investigation of the mechanical properties of interlayer interfaces therefore is of utmost importance.

While mechanical properties are a vast playground for theoretical modelling, microscopic size of the flakes and heterostructures leaves not so much room to access these properties experimentally. Indeed standard mechanical probes applicable to membranes require a rather large lateral size or sophisticated nano-fabrication\cite{mechreview1, mechreview2}.  
Raman scattering is a powerful tool, however due to selection rules it says almost nothing about the low-frequency acoustical phonons.
Nanoindentation\cite{gao2015} and atomic force microscopy\cite{falin2021} do  not provide a comprehensive information on mechanical properties of multilayer structure as they require subtraction of huge contribution related to the substrate. 
Picosecond ultrasonics or picoacoustics appears to be much more effective tool to access the rigidity of the flakes and interfaces in the case of thin multilayer structures\cite{reviewCP2020}.

Acoustical imaging is  well developed for standard semiconducting heterostructures\cite{Klokov2020}.
 Thin layered materials and Van der Waals heterostructures remain insufficiently explored. In Refs.\cite{greener2018,greener2019} several picoacoustical approaches were first applied to thin flakes and Van der Waals heterostructures. In particular the bubbles were identified using the acoustical resonance technique. It was demonstrated that the acoustical coupling of the layers and substrate fluctuates strongly, that impedes comprehensive studies of mechanical parameters of layers and interfaces.  

In this paper we substantially improve the sensitivity of the acoustical probing of Van der Waals heterostructures. 
We use Al film as a transducer that effectively converts essential part of the pump laser pulse energy into the coherent acoustical pulse. 
We also improve the sensitivity of the scheme using the Sagnak interferometer detection of the variation of the reflection coefficient\cite{Tashizaki}. 
The abovementioned improvements allowed us to detect both spectrally narrow high- and low-frequency Fourier components of the picoacoustical response and hence to perform a tomography of Van der Waals heterostructures.
Our measurements allow to find the acoustic parameters of the layers and coupling acoustical constants from the modelling of the time-resolved responses. 

The insulated layers of h-BN and WSe$_2$ were exfoliated on oxidized Si using the gold-mediated method\cite{liu2020}. The thickness of the flakes was determined using AFM. 
Van der Waals heterostructures out of these flakes were then assembled on Al$_2$O$_3$ substrate with a preliminary lithographically defined\cite{pugachev2021} array of binary markers. We  
using dry hot pick up techinique\cite{pizzochero2016} with a home-made transfer machine\cite{martanov2020} at ambient conditions.
Then $\sim$30 nm Al layer was e-beam evaporated atop. {The details of sample fabrication and characterization are given in Supplementary materials.}

{An optical pump-probe scheme, similar to Ref.\cite{Tashizaki} (see Supplementary information) allows to perform time- and spatially-resolved measurements of the reflectivity coefficient. 
A pump pulse with $\sim400$ nm wavelength }generates a picosecond elastic pulse in aluminum propagating perpendicular to the plane direction and partially reflecting from the heterostructural interfaces. 
The propagation of elastic pulses in the structure leads to a change in its reflection coefficient, which is detected by a {$\sim 800$ nm wavelength }probe pulse. For the subsequent analysis the temporal variation of the reflection coefficient after the subtraction of the slow component is expanded into Fourier spectrum. An ideal coupling (acoustic mismatching) means that the acoustical displacement of two neighboring points belonging to different materials is the same. In most of cases however, especially relevant to layered materials, one has to introduce a non-ideal coupling that can be modeled as a mass-less spring with rigidity uniformly distributed over area. 
The rigidity of this spring is one of the key mechanical parameters of the Van der Waals interface. Nevertheless so far this parameter for the structures under study was not accessed experimentally.


To model the responces of the structures the strain pulse excitation and propagation through a sequence of the layers on semi-infinite substrate was calculated as in Ref\cite{Klokov2020}. Then optical response of the structure was obtained anolougous to Ref. \cite{Perrin}.

\begin{figure}[h!]
    \centering
    \includegraphics[width=0.5\textwidth
    ]{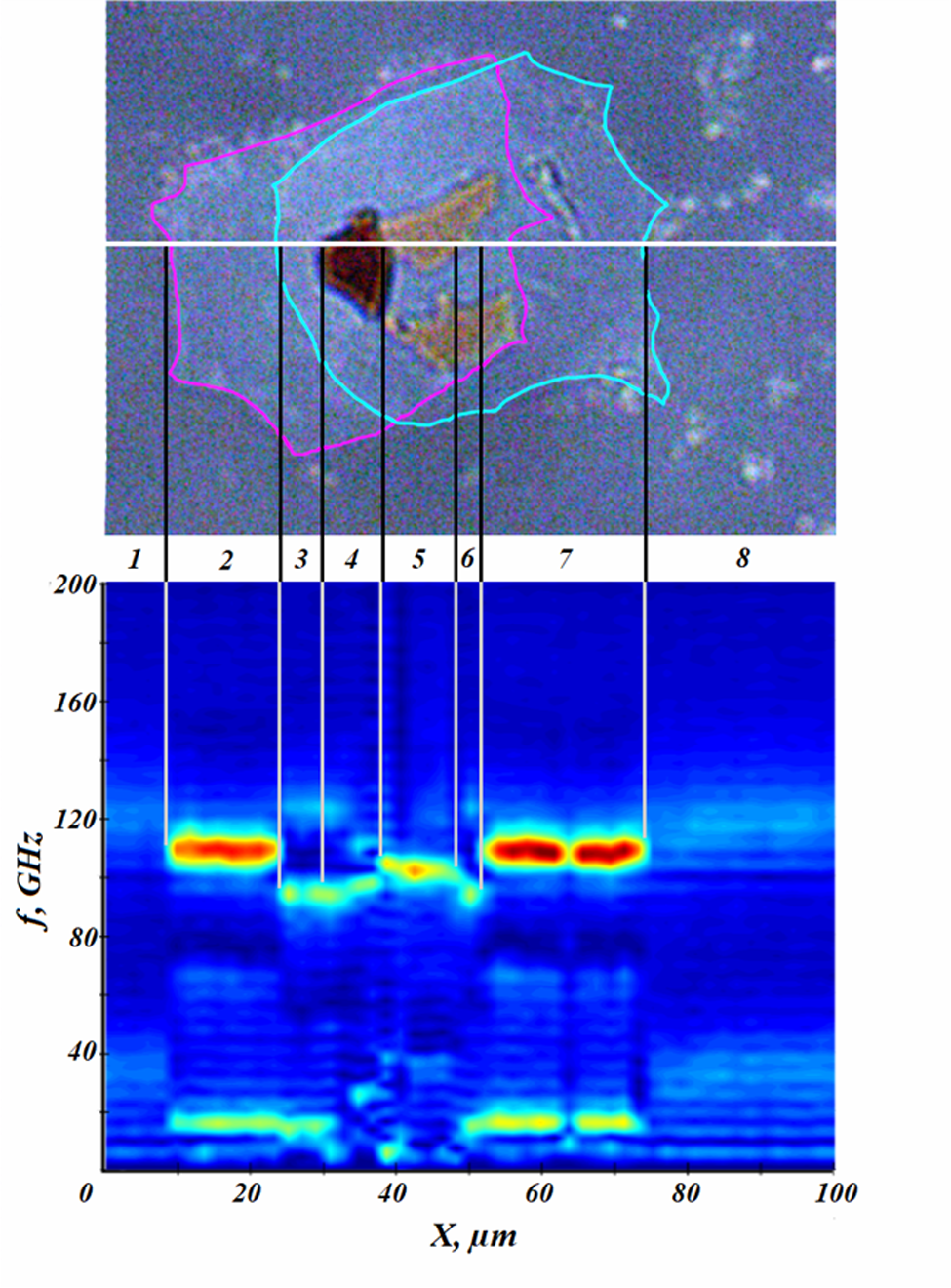}
    \caption{A photo of the sample with the scan line and a color plot of the picoacoustical reflectivity spectra as function of the the position along the line. The regions with different sequence of layers are identified.}
    \label{spectrscan}
\end{figure}

In order to demonstrate a sensitivity of the picoacoustical technique to the composition of the heterostructure we show in Figure \ref{spectrscan} a photo of the heterostructure and a color plot of the spectrum, scanned along the straight line. { We distinctly resolve} 
different composition regions 
in the acoustical response. 
For example, in the domains 2 and 7, where only one 10 nm thick hBN layer is located, the spectrum consists of two intensive lines at $\sim$110 and $\sim$16 GHz, respectively and a lower intensity line at $\sim$70 GHz. In the domains 3 and 6 , where a double layer of hBN is located, the high frequency peak splits into two and the frequency of the other two peaks decreases. In the domains with WSe$_2$ (5 and 20 nm thickness) the low frequency line disappears, and a new line emerge at $40-50$ GHz.

\begin{figure}[t]
    \leftline{\includegraphics[width=\linewidth]{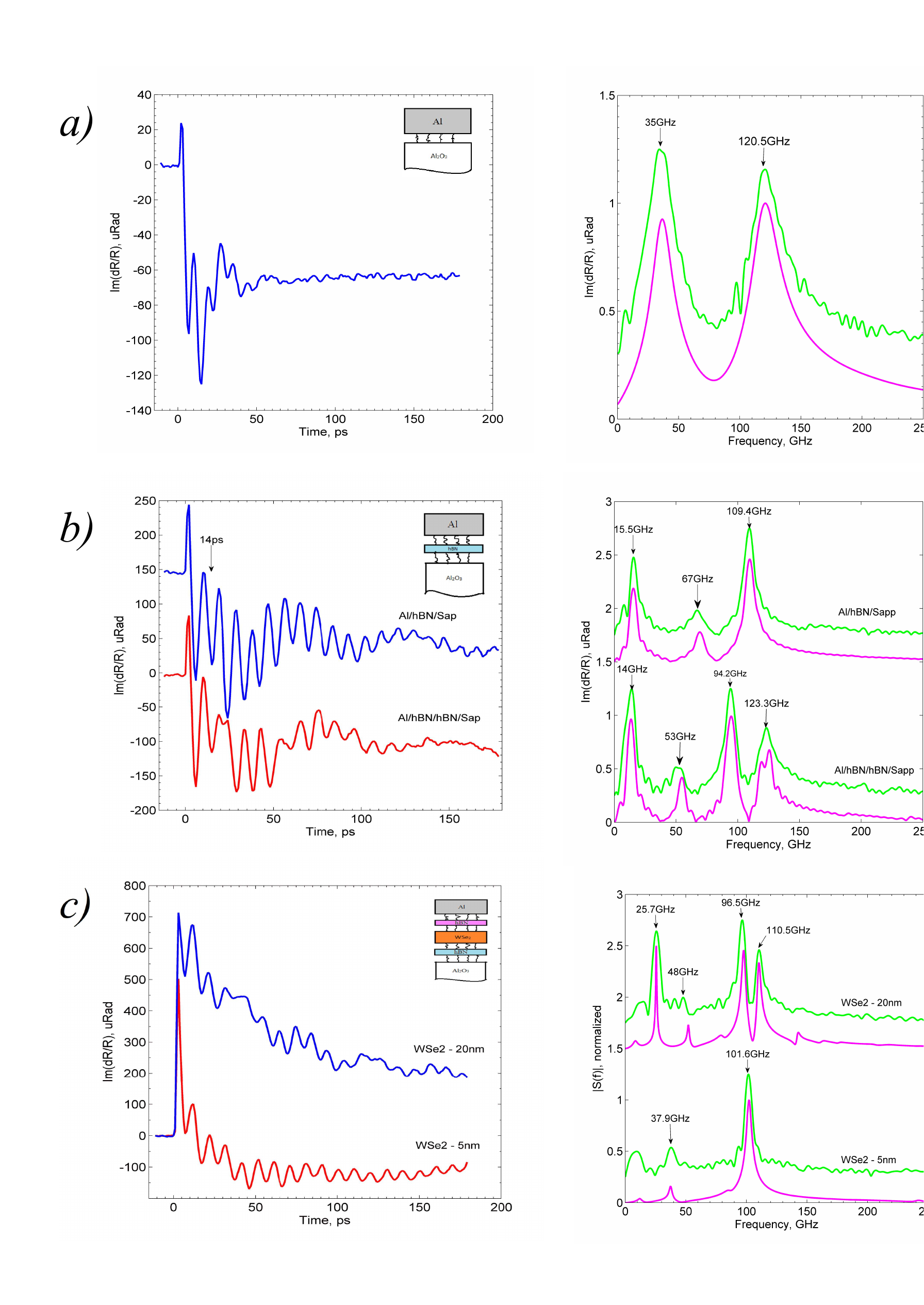}}
    \caption{    Time resolved response (left) and spectra (right; green - experiment, magenta - simulation) collected from different regions in Fig.\ref{spectrscan}: a) regions 1 and 8 ; b) region 2 (blue) and 3 (red) c) region 4 (blue) and 5 (red). The insets show sample structure.}
    \label{responses}
\end{figure}

To 
obtain acoustical parameters of the layers and interfaces in Fig.\ref{responses} we consider time dependencies of the reflectivity collected at various representative points of the stack and compare the experimental Fourier spectra with the calculated ones. By adding new layers we specified the unknown parameters, such as photothermal and photoelastic coefficients. 

Far away from the heterostructure (domains 1 and 8 in Fig. \ref{spectrscan}) the response is determined by Al film solely and its interface with Al$_2$O$_3$ substrate. As seen in Fig. \ref{responses}a just after the laser pulse a negative oscillating response is observed. Then decaying oscillations develop atop of a smooth dependence.


A Fourier transform of the observed pattern clearly produces two wide peaks with relevant frequencies 39 GHz and 117 GHz, respectively. {Even for the case of sapphire and Al an acoustic matching can not account for the observed spectrum, in order to model the structure one has to assume that even Al to Al$_2$O$_3$ interface bonds are not ideal and have a finite rigidity $\sim9\cdot10^{18}$ N/m$^3$.  }

Physically presence of two spectral lines in Fig.\ref{responses}a 
means that the Al layer oscillates as a whole (low frequency peak) and that there is a travelling and reflected wave (high frequency peak).


Fig \ref{responses}a {(right panel)} shows a simulated spectrum of the optical response that assumes all Al and laser pulse parameters are known\cite{Klokov2020}, Al thickness (about 30~nm), rigidity of the interface and the viscoelastic hypersound damping are adjustable parameters. The agreement of the modeled thickness and the value 35~nm indicated by the quartz thickness monitor during the evaporation indicates for an adequacy of our model.

Adding up $\sim$10 nm hBN layer under the Al transducer (blue line in Fig. \ref{responses}b) apparently makes spectrum more complicated, because reflections from both Al/hBN and hBN/sapphire interfaces produce additional eigen frequencies. Markedly the spectral lines become narrower, because hBN and Al$_2$O$_3$ are not well matched and sound reflects more effectively from both interfaces.  For example 109 GHz line has a 7.3GHz width that corresponds to Q=15.

The simulation with the known acoustic parameters of the Al film produces a brilliant agreement of the spectrum and allows to find the  Al/ hBN and hBN/sapphire coupling constants having the thickness of Al layer known.

{Noteworthy we could but do not need to introduce inelastic scattering of the sound wave in the layered crystals and at the interfaces. This means that such scattering physically exists only in amorphous Al film.}

In Fig. \ref{responses}b {by red line }we show the response for the domain where a stack of two 10 nm thick hBN layers
is located under the Al layer.
The time domain signals from single- and two-layer hBN during the first two periods of oscillations are pretty much the same, because they are determined by physically the same process - reflection of the elastic pulse from the Al/hBN interface. 
In this system we observe almost ideal acoustic matching of two hBN layers. The high frequency peak splits because two oscillators ({Al resonator and hBN} resonator) have close frequencies {(see simulation below)}.

 In Fig. \ref{responses}c {$\sim$5 and $\sim$}20 nm thick layers of heavy material WSe$_2$ are added between the hBN layers. An indicator of WSe$_2$-layer is a slow dynamics atop of oscillations. This decay is related to the change of the WSe$_2$ complex refraction index due to the photoexcited carriers, as a small part ($<5$\%) of the optical pump and probe pulses passes through the Al film. The acoustical mismatch with the other layers makes reflection of the hypersound from the hBN/WSe$_2$ interface even more effective, {that leads to further narrowing of the spectral lines.} A simulation allows to find the parameters of all the interfaces.

\begin{figure}[h]
    \centering
    \includegraphics[width=0.5\textwidth]{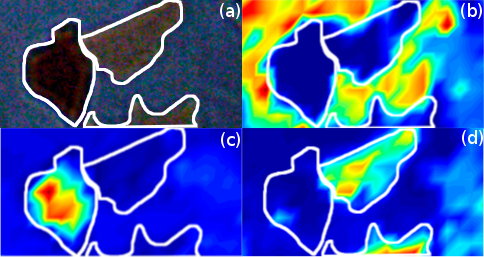}
    \caption{Optical image of the sample(a) and spatial maps of the  Fourier component at three different frequencies: (b) 14 GHz; (c)26 GHZ; (d)101 GHz.}
    \label{scans}
\end{figure}

   In {Supplementary materials we summarize acoustical parameters of the interfaces found within these fits. }
Mapping of some of Fourier harmonics over the sample may serve as a {compositional tomography tool}. In Fig. \ref{scans}a-d we plot an optical image of the h-BN/WSe$_2$/h-BN sample and the corresponding maps of the Fourier components at various frequencies.


\begin{figure}[h]
    \centering
    \includegraphics[width=\linewidth]{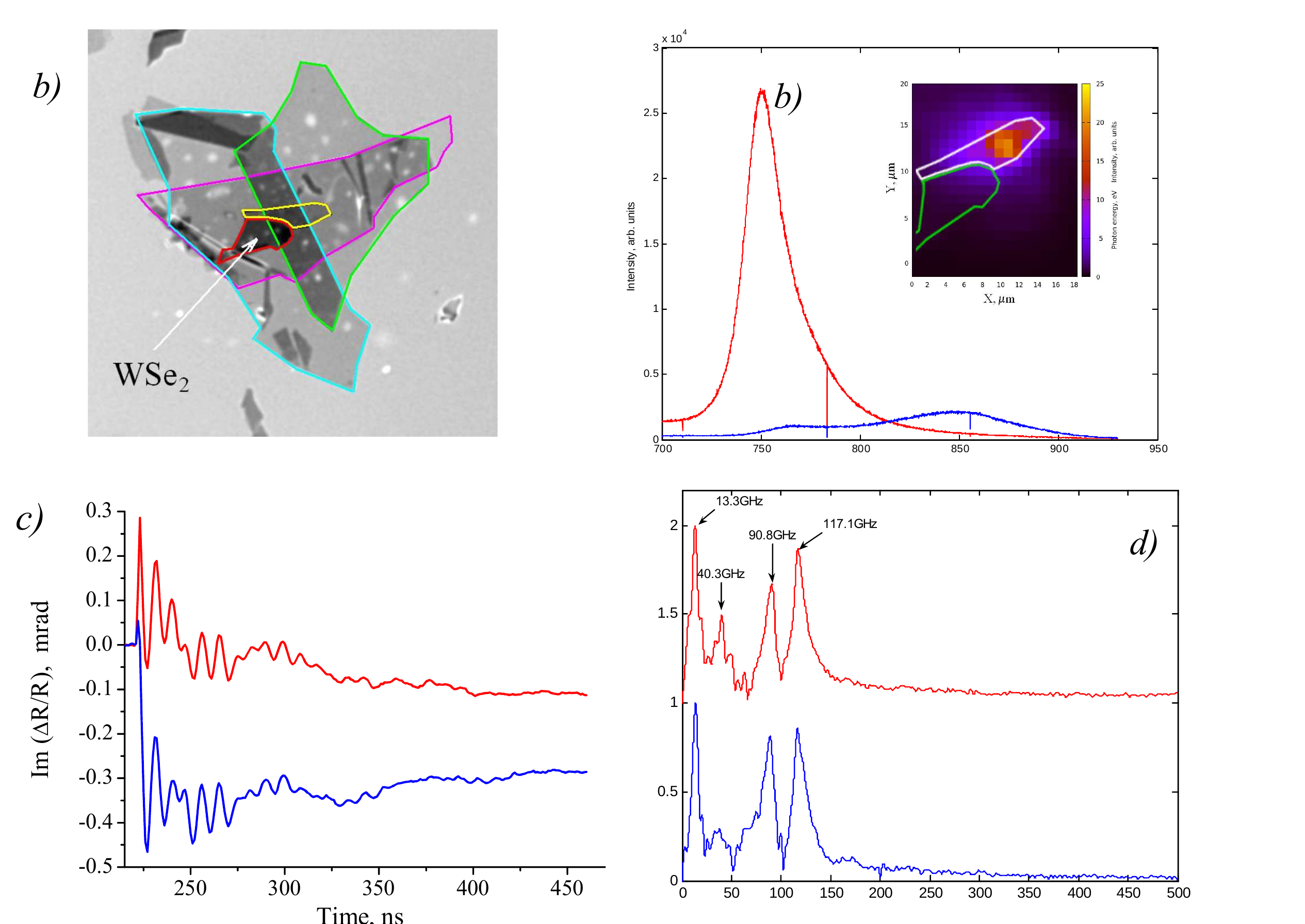}
    \caption{ Image of the sample(a) and photoluminescence spectra from mono- and multilayer segments at 300K. Insert shows spatial maps of the photoluminescence  at 738 nm$<\lambda<$760 nm after the 472 nm excitation. (c) and (d) are reflectivity photoresponse and its fourier transform from the hBN-monolayer WSe$_2$-hBN (red) and hBN-hBN (blue) segments, respectively.}
    \label{monolayer}
\end{figure}

{To answer the question whether this optical technique provides sensitivity to monolayers we fabricated a heterostructure shown in Fig.\ref{monolayer}a. Measurements of Raman scattering and photoluminescence mapping Fig.\ref{monolayer}b, analogous to Ref. \cite{akmaev2020} (see Supplementary information for detail) allows us to clearly identify a single monolayer domain. Photoresponse of the reflectivity for two neighbouring areas (with and without a monolayer) is shown in Fig.\ref{monolayer}c, and the corresponding spectra are shown in Fig.\ref{monolayer}d.
A characteristic and reproducible spectral feature of the embedded WSe$_2$ monolayer is a formation of $\sim$40 GHz peak. Thus the presented pico-acoustic technique demonstrates a monolayer sensitivity.}

\begin{figure}[h]
\centering
\includegraphics[scale=0.4]{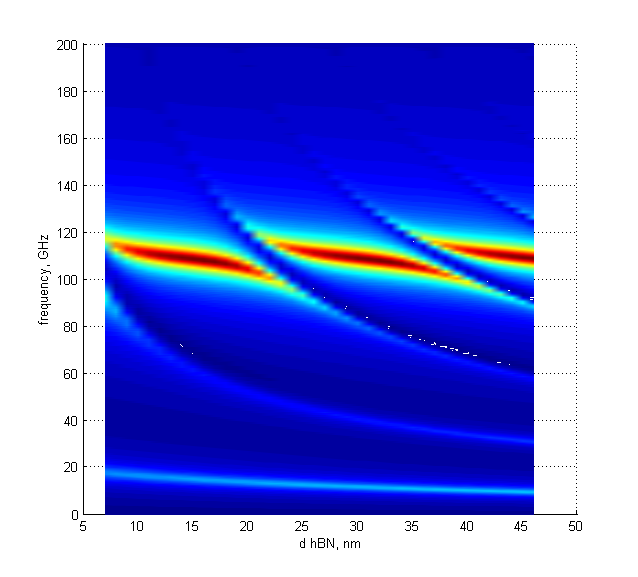}
\caption{Simulation of the picoacoustical response spectrum of hBN layer under 30 nm Al transducer on sapphire substrate as function of hBN thickness with the experimentally known acoustic couplings at the interfaces.}
\label{simulation}
\end{figure}

{\bf Discussion.} 
We show in Fig. \ref{simulation} a simulation of the Fourier spectra of Al/hBN/sapphire structure with the interface parameters determined from the experimental data as a function of hBN width. Different types of spectral lines can be clearly identified: 
\begin{itemize}
    \item Low-frequency peak, that is sensitive to the mass per unit area of hBN with Al transducer and rigidity of hBN-to-substrate bond. The heavier the structure the lower this frequency is. Exactly this peak corresponds to the lowest frequency in the bottom panel of Fig. \ref{spectrscan}.
    \item The most intensive pundamental peak of Al transducer ($\sim$110 Ghz in our case). 
    \item Fundamental peaks of hBN+Al, those frequencies strongly drop with thickness of hBN layer. At frequencies $>$110 Ghz this mode corresponds to hBN solely.
\end{itemize}
In the domains where peaks of Al and hBN become close to each other peak anticrossing is observed. In the case of multilayer heterostructures peaks of combined nature might emerge depending on the strength of the acoustic coupling between layers and their thicknesses.

The hypersonic response of the structures from Ref.\cite{greener2018} (Figs. 6 and 8) with Al transducer seems to be similar to our structures, shown in Fig. \ref{responses}b,d, because the thicknesses of Al and layered materials are close. Nevertheless the linewidths in our case are several times smaller:{ our structures are much more effective acoustical resonators and therefore allow to extract a bunch of information about the interfaces}. There are two main differences between Ref.\cite{greener2018} and this work:
\begin{itemize}
    \item Fabrication process. We believe the key element is the thermal treatment of the heterostructures.
    \item Experimental method. Interferometric technique based on Sagnac interferometer rules out the low-frequency noise components and guarantees high quality of the signal and the Fourier transformation. We also use a small spot size that prevents from unwanted averaging over too much area.
    In this paper we fitted time-resolved responses of the heterostructures, rather than the Fourier spectra. This is important, because the positions of the spectral lines depend not only on eigen frequencies but also on the sensitivity function \cite{Matsuda2015}, that relates optical response with the strain distribution in the sample. 

    \end{itemize}

Let us discuss the future development and applications of the presented picoacoustical technique.
First of all, a systematic study is needed to reveal a role of interface preparation and twist angle on the acoustical properties of the interfaces. The technique would be highly demanded by technology, since many efforts are performed to make separate layers and Van der Waals heterostructures uniform for the device applications, in particular to fight against bubbles and ripples.

Demonstrated increase of the acoustical Q-factor in the Van der Waals heterostructures makes them especially promising for amplification of the acoustic field. Besides purely hypersonic experiments, these resonators could be a playground for interactions of the acoustic pulse and electronic system. For example ultrafast acoustical field can be used to modulate emission in atomically thin materials and heterostructures, similarly to epitaxial quantum wells \cite{scherbakov2008}. We would expect that intensive picosecond strain pulse will change the reflectivity and exciton wavelength in monolayers, thus making TMDC potentially suitable for acousto-optical modulation applications. Placement of the piezoelectric\cite{cui2018} 2D layers may force formation of electric field, and hence sub-THz radiation emission and numerous other effects on the electronic system. Another application of the acoustical resonators is the old-standing problem of the nonequilibrium amplification of superconductivity.


{\bf Conclusions.} 
{In this paper we study propagation of picosecond elastic pulse in van der Waals heterostructures using an optical pump-probe technique with spatial resolution.  A combination of top-evaporated Al transducer and interferometric detection technique allowed to perform highly sensitive 3D hypersonic microscopy, and, in particular, demonstrate WSe$_2$ monolayers embedded in hBN. 
Approximation of the photoresponse, collected from different part of heterostructures allowed to determine rigidity of the Van der Waals bonds  at the hBN/Sapphire, hBN/Al, hBN/hBN and hBN/WSe$_2$ interfaces. Our results indicate that the diffuse scattering of the sub-150 Ghz phonons inside the layers and at the interfaces is negligible. This implies a great potential of van-der Waals heterostructures for the nanoacoustic applications.} 


{\bf Acknowledgements.} The work was supported by the Government of the Russian Federation (Contract No. 075-15-2021-598 at the P.N. Lebedev Physical Institute) and Russian Foundation for Basic Research (Project no. 19-02-00952a).

\end{document}